# Trapping of solute atoms at grain boundaries in GdNi2

Ryan Murray, Debashis Banerjee[*], and Gary S. Collins

*Department of Physics and Astronomy, Washington State University, Pullman, WA 99164, USA*

## Abstract

Lattice locations of $^{111}$In impurity probe atoms in intermetallic GdNi$_2$ were studied as a function of alloy composition and temperature using perturbed angular correlation spectroscopy (PAC). Three nuclear quadrupole interaction signals were detected and their equilibrium site fractions were measured up to 700 $^{\circ}$C.  Two signals have well-defined electric field gradients (EFGs) and are attributed to In-probes on Gd- and Ni-sites in a well-ordered lattice.   A third, inhomogeneously broadened signal was observed at low temperature.  This is attributed to trapping, or segregation, of In-probes to lattice sinks such as grain boundaries (GB) that have a large multiplicity of local environments and EFGs.  Changes in site fractions were *reversible* above 300$^{\circ}$C.  Measurements were made on a pair of samples that were richer and poorer in Gd.  Remarkably, the GB-site was populated only in the more Gd-rich sample.  This is explained by the hypothesis that excess Gd segregates to the grain boundaries and provides a lower enthalpy environment for In-probe atoms.  The observations are discussed in relation to a three-level quantum system.  Enthalpy differences between the levels were determined from measurements of temperature dependences of ratios of site fractions.  The enthalpy of transfer of In-probes from the Gd- to Ni-sublattice was found to be much smaller in the Gd-rich sample.  This is attributed to a large temperature-dependence in the degeneracies of levels available to In-solutes in the phase, leading to an effective transfer enthalpy that differs greatly from the difference in site-enthalpies.   A possible scenario is discussed.   Different segregation enthalpies were measured for In-solute transferring from GB sites to Gd- and Ni-sites, whereas only an average value can be determined through macroscopic measurements.  Additional measurements on GdNi$_2$ alloys doped with 1-2 at.% Cu are also presented and discussed.

---

[*] Permanent address:  Accelerator Chemistry Section, RCD(BARC), Variable Energy Cyclotron Centre, Kolkata 700064, India

## Introduction

A subject of considerable interest in the study of solid solutions is the site-preference of impurity atoms [see, e.g., 1]. In a well-ordered compound, impurities may occupy any of the normal crystallographic sites, with different enthalpies for each site. Impurities may also interact with point defects or locate in extended defects such as dislocations and grain boundaries. Extended defects in general have large multiplicities of local environments. This situation lends itself well to investigation through measurement of nuclear quadrupole interactions at nuclei of the impurity species. Impurity probes on crystallographic sites in well-ordered compounds experience essentially unique electric field gradients (EFGs) whereas probes in "lattice sink" sites more likely experience highly inhomogeneous EFGs.

The present work examines lattice locations of In-probe atoms in the compound $GdNi_2$. $GdNi_2$ has the cubic Laves $Cu_2Mg$ structure, with Gd- and Ni-sites in the perfect crystal having $\bar{4}3m$ (cubic) and $\bar{3}m$ (axial) point symmetries, respectively [2]. The cubic site has zero electric field gradient (EFG) and nuclei at that site will therefore have zero quadrupole interaction frequency. The axial site has a nonzero EFG and nuclei there will have a finite quadrupole interaction frequency. This makes it easy to determine site-fractions of impurity probe atoms on the two sublattices. An earlier study was made of isostructural $GdAl_2$ [3]. The behavior found there was that of a two-level quantum system, with a "transfer enthalpy" determined by measuring the temperature dependence of the ratio of site-fractions of probes on the two sites. The enthalpy difference found was 0.343(3) eV, independent of composition, with the site-fraction of probes on Al-sites increasing with increasing temperature.

In the present work, measurements were made in a similar way on two $GdNi_2$ samples having mean compositions 33.5 and 34.9 at.% Gd. For the more Gd-poor sample, two unbroadened signals analogous to those observed in $GdAl_2$ were observed. The transfer enthalpy of In-probes between the Gd- and Ni-sublattices will be shown below to be 0.47(10) eV, somewhat larger than the 0.343 eV value measured in $GdAl_2$. For the more Gd-rich sample, those two signals were observed as well as an additional signal having a large amount of inhomogeneous broadening. Because this signal was observed at the same time as the well-defined signals for isolated probes on Gd- and Ni-sites, it is concluded that the broadened signal arises from probes in lattice sinks such as dislocations or grain boundaries. Since all three signals were observed after lengthy measurements at elevated temperatures, during which time dislocations should largely anneal out, the lattice sinks are identified with grain boundaries. Site fraction changes for all three signals were reversible above 300 °C, and quenched below.

The question arises why indium solutes segregate to GBs only in the more Gd-rich sample. This is attributed to a difference in the nature of the GBs. Highly ordered phases such as $GdNi_2$ appear as "line compounds" in binary phase diagrams. While this suggests that they have a definite composition, perhaps at a stoichiometric composition, there always exists a small but finite width to the phase field [4]. The width arises from small concentrations of intrinsic point defects such as antisite atoms or vacancies. An excess of Gd-atoms beyond the boundary composition might be accommodated in small volumes of a more Gd-rich phase or in part by segregation of some Gd to grain boundaries. It is our hypothesis that some excess Gd segregates to grain boundaries in the Gd-rich sample and then In-probe atoms segregate to the Gd-decorated grain boundaries owing to a favorable energy of solution.

Measurements were also made on samples doped with 1-2 at.% Cu at Gd-rich and Gd-poor compositions. Just as in the undoped samples, it was found that In-probes segregate at low temperature to GB sites in samples that were Gd-rich but not Gd-poor. Measurements also showed that in Gd-poor samples, Cu-solutes formed bound states with In-probes on Gd-sites at low temperature.

## Experimental

Samples were doped with $^{111}$In activity (mole fraction $\sim 10^{-11}$) by melting alloy constituents and activity together under argon in a small arc-furnace. Metal purities were 3N for Gd and 4N for Ni, and the $^{111}$In activity was carrier-free. Two $GdNi_2$ samples were made having Gd-poor and Gd-rich mean compositions, $Gd_{33.5(1.5)}Ni_{66.5(1.5)}$ and $Gd_{34.9(0.6)}Ni_{65.1(0.6)}$, with uncertainties reflecting the precision of measured masses of starting materials and taking into account any mass loss during melting. The study of Gd-poor and Gd-rich samples was expected to lead to greater probe occupation of one sublattice or the other according to a heuristic rule that impurity atoms tend to occupy the sublattice of an element in which there is a deficiency [3]. In addition, two samples doped with Cu were made having mean compositions $Gd_{33.1(2)}Ni_{64.9(2)}Cu_{2.0(2)}$ and $Gd_{34.5(1)}Ni_{64.5(1)}Cu_{1.0(1)}$.

$^{111}$In decays to the second excited state of $^{111}$Cd by electron-capture with a mean life of 4.0 days. $^{111}$Cd subsequently decays to the ground state with successive emission of 173 and 247 keV gamma rays, the intermediate 247 keV PAC level having a mean life of 120 ns. PAC measurements were made using four-detector PAC spectrometers employing 1.5 x 1.0 inch $BaF_2$ scintillators. Coincidences detected between the 173 and 247 keV gamma rays give lifetimes of the individual decays of the PAC level that were accumulated over a time of one day. For each measurement, four time-coincidence spectra were simultaneously accumulated, two each at detector angles of 180° and 90° relative to the sample. After fitting and subtracting accidental backgrounds, spectra were geometrically averaged and then combined to yield the experimental "PAC spectrum". The PAC spectrum was fitted with a superposition of perturbation functions for the nuclear quadrupole interaction of a spin I=5/2 level. For each site occupied by probe atoms, the perturbation function is given by:

$$G_2(t) = s_0 + \sum_{n=1}^{3} s_n \cos(\omega_n t) \exp\left(-\frac{\omega_n^2}{\omega_1^2}\frac{\sigma^2 t^2}{2}\right). \tag{1}$$

The three frequency components arise from splitting of the spin $I=5/2$ PAC level in an EFG. They have the properties that $\omega_1 < \omega_2 < 2\omega_1$ and that $\omega_3 = \omega_1 + \omega_2$. The three frequencies depend on the quadrupole interaction strength and EFG asymmetry parameter $\eta \equiv \left|\frac{V_{xx} - V_{yy}}{V_{zz}}\right|$, in which $V_{zz}$ is the principal component of the EFG tensor. The fundamental observed frequency, $\omega_1$, is a function of $V_{zz}$ and the asymmetry parameter. For the special case of axial symmetry, $\eta = 0$, $\omega_1 : \omega_2 : \omega_3 = 1 : 2 : 3$, and $\omega_1 = \left| \frac{3\pi}{10} \frac{eQV_{zz}}{h} \right|$, in which $Q$ is the quadrupole moment of the PAC level. For each signal, the amplitudes sum to unity, $\sum_{n=0}^{3} s_n = 1$. The overall amplitude of each fitted perturbation function is equal to the site fraction for that signal. The parameter $\sigma$ describes the width of the frequency distribution of a signal caused by weak EFG disturbances from distant defects. $\omega_1$ is used to label the observed signals below. For the highly broadened GB signal, fitting was constrained using an empirical value $\eta = 0.6$ that has previously been shown to give satisfactory fits when the fractional frequency broadening is large [5]. For an overview of PAC spectroscopy and methodology applied to solids, see [6].

## Results

PAC spectra are shown for pairs of GdNi$_2$ and GdNi$_2$(Cu) samples in Figs. 1 and 2, with the more Gd-rich spectra shown on the right. (The data at apparent negative times in the figures are independent data that "mirror" the positive delayed-coincidence time spectrum, and are obtained by gating times of arrival of the 173 and 247 keV gammas in reverse order. Such "double-sided" spectra are preferred because they make clearer the shape of the spectrum near time zero.)

**Signals observed.**

Representative spectra are shown in Figs. 1 and 2. Two signals were observed at high temperature for all samples. In all spectra, a zero-frequency offset was observed and attributed to probe atoms on cubic Gd-sites, having a fitted site fraction of 50-80%. A high-frequency axially-symmetric quadrupole interaction with 40-ns period was observed having a site fraction up to 20% and is attributed to In-probes on Ni-sites.

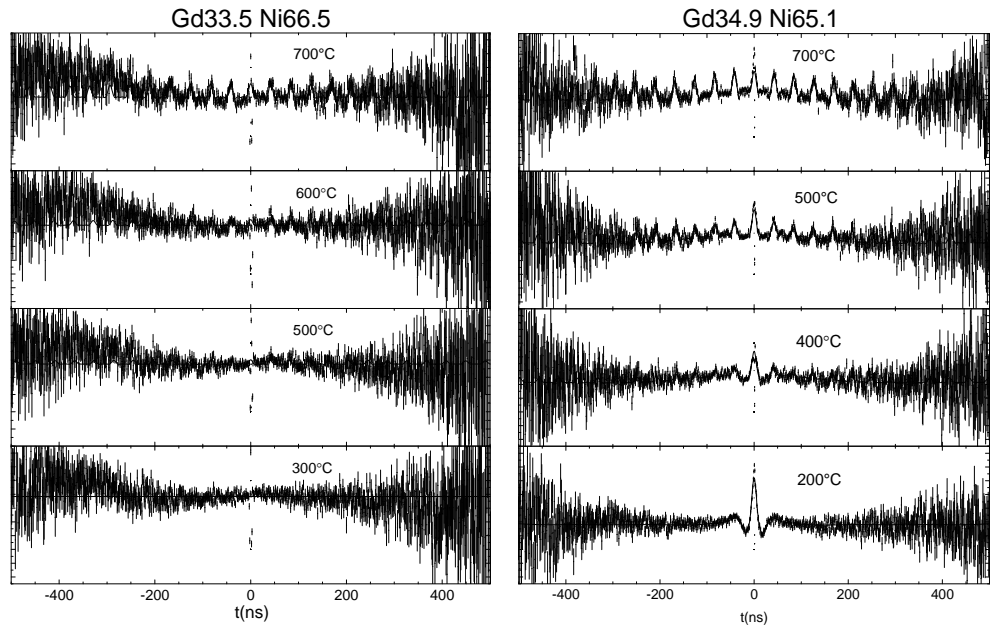

Figure 1. PAC spectra of GdNi2 measured at the indicated temperatures. <u>Left</u>: Gd33.5 Ni66.5. <u>Right</u>: Gd34.9 Ni65.1.

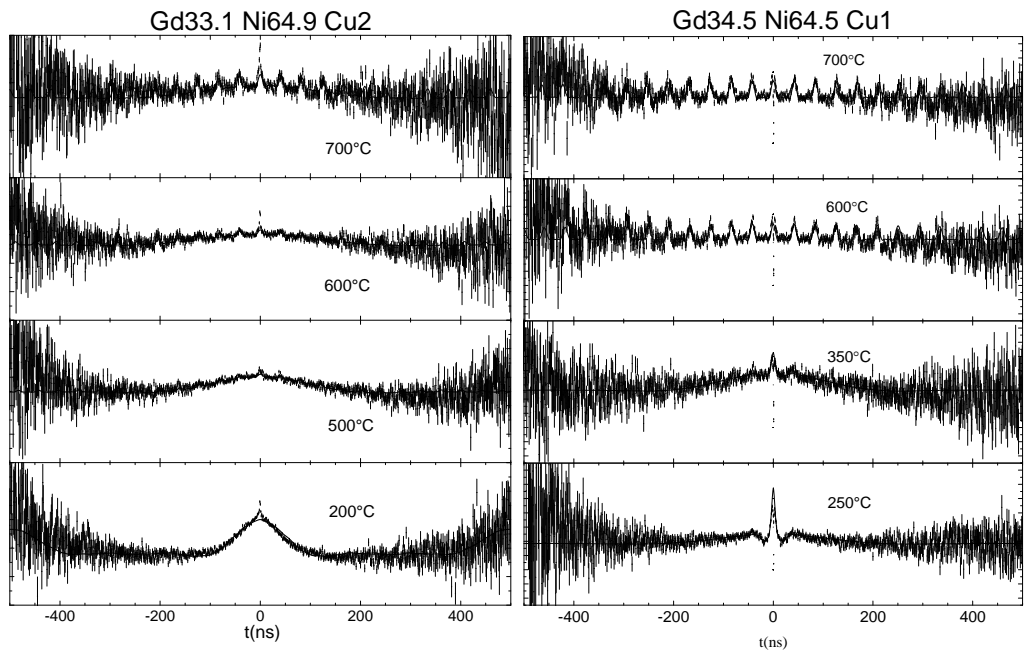

Figure 2. PAC spectra of Cu-doped GdNi$_2$ measured at the indicated temperatures. <u>Left</u>: Gd33 Ni65 Cu2. <u>Right</u>: Gd35 Ni65 Cu1.

As the temperature increases, the site-fraction of the high frequency signal increases. However, at low temperature in the undoped Gd-poor sample (Fig. 1, left), the high-frequency interaction for the Ni-site is unobserved. For both Gd-rich samples (right hand sides of Figs. 1 and 2), as temperature decreases, there is an increase in the amplitude of an inhomogeneously broadened signal that is visible in the spectra as a "glitch" within ±50 ns of time zero.

The spectrum of the Gd-poor Cu-doped sample at 200 °C (Fig. 2, left) exhibits an additional low-frequency interaction of 15 Mrad/s that is not observed at higher temperature or in the undoped sample. This is attributed to an attractive interaction forming pairs of close neighbor Cu-solutes and In-probes, both on Gd-sites.

Table 1 summarizes all observed signals and their attributions. Differences between fitted parameters for the GB signal in the undoped and Cu-doped samples are not considered to be significant.

Table 1. Quadrupole interaction parameters $\omega_1$, $\eta$ and $\sigma$ measured at the indicated temperatures. Column 5 indicates samples in which the signal is observed. Column 6 gives attributions of the signals to sites. The difference between the two listed GB interactions is not considered significant.

| $\omega_1$ (Mrad/s) | $\eta$ | $\sigma$ (Mrad/s) | T | Samples | Attribution |
|---|---|---|---|---|---|
| 148.7(2) | ~0 | <2 | 700 °C | All | $In_{Ni}$ |
| 0 | ~0 | <2 | - | All | $In_{Gd}$ |
| 138 | 0.6 | 58 | 200 °C | Gd-rich only, Cu-doped | $In_{GB}$ |
| 151 | 0.6 | 124 | 200 °C | Gd-rich only, undoped | $In_{GB}$ |
| 10.7(2) | ~0 | 9.5(3) | 200 °C | Gd-poor only, Cu-doped | $In_{Gd} + Cu_{Gd}$ |

## Temperature dependences of site fractions.

Figure 3 shows site fractions plotted as a function of temperature for the undoped Gd-poor and Gd-rich samples (cf. spectra shown in Fig. 1). For the Gd-poor sample (left), the observed trend as the temperature increases is a transfer of In-probes from Gd-sites to Ni-sites. For the Gd-rich sample (right), there is a transfer with increasing temperature of probes from GB sites to both Gd- and Ni- sites.

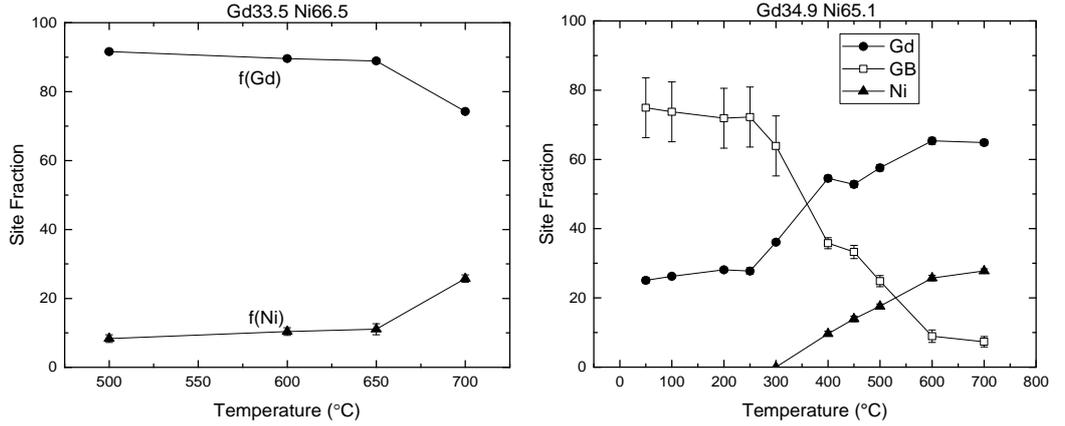

Figure 3. Temperature dependences of site fractions of In-solutes in the undoped samples. Left: Gd33.5 Ni66.5. Right: Gd34.9 Ni65.1. The sites include In-probes on crystallographic Gd and Ni sublattices and in a grain-boundary environment (GB). Measurements below 300°C are on quenched samples.

## Solute transfer between Gd- and Ni- sites.

Site fractions of probe atoms are in principle governed by the differences in enthalpies of probes at the two sites. In thermal equilibrium, the ratio of site fractions of an impurity on two sites labeled 1 and 2 can be written as

$$\frac{f_2}{f_1} = \exp((S_2 - S_1)/k_B)\exp((H_2 - H_1)/k_B T) = \exp(\Delta S/k_B)\exp(Q/k_B T), \quad (2)$$

in which $S_n$ is the vibrational entropy of the impurity on site $n$, $H_n$ is the enthalpy of the impurity on site $n$, and $Q=H_2-H_1$ is the difference in site-enthalpies. Fig. 4 shows an Arrhenius plot of the ratio of site fractions of In-probes on Gd and Ni sites for the undoped Gd-poor and Gd-rich samples. For the 33.5% sample (for which the GB signal was not observed), the transfer enthalpy between Gd- and Ni-sites was fitted as 0.47(13) eV, which is marginally consistent with the value 0.343(3) eV observed for GdAl$_2$. (The larger uncertainty for GdNi$_2$ is due to the small site fraction for probes on Ni-sites; cf. spectra in Fig. 3, left).

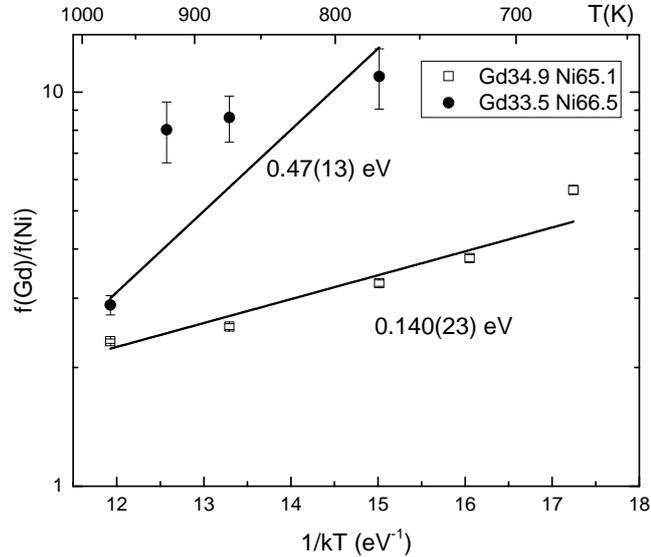

Figure 4. Arrhenius plot of ratios of site fractions of probe atoms on Gd- and Ni-sites in the two undoped samples.

For the Gd34.9 sample, with Gd-, Ni- and GB sites, the observed transfer enthalpy between Gd- and Ni-sites was much smaller, 0.14 eV. This large difference is surprising in view of Eq. 2 because site-enthalpies of atoms in solids depend almost entirely on the local atomic environments, perhaps out to the second atomic shell surrounding a site. Since the Gd- and Ni-sites have well-defined local environments, one expects the experimental Q to be a constant.

## Segregation enthalpies for In-solutes.

Fig. 5 shows transfer enthalpies of probes between GB and Gd sites, and between GB and Ni sites, for the Gd34.9 sample, plotted in the same way as Fig. 4. These are customarily termed segregation enthalpies.

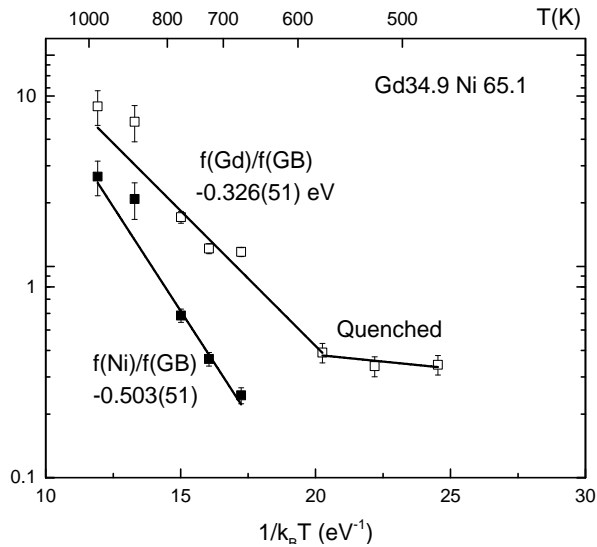

Figure 5. Arrhenius plot of the ratios of site fractions of In-probes on Gd or Ni sites to those on GB-sites for the 34.9 at.% Gd sample. (At and below about 600K, the sample does not reach thermal equilibrium in a time of the order of one day.)

It can be seen that In probes transfer out of GB sites to Gd- and Ni-sites as temperature increases, but with activation enthalpies, or segregation enthalpies, that *differ* for Gd-sites (0.33 eV) and Ni-sites (0.50 eV). A segregation enthalpy measured using a macroscopic method that averages over all crystal sites would instead yield a single value that would be a kind of average over the different crystallographic sites. The two segregation enthalpies and the transfer enthalpy between Gd- and Ni-sites obey a sum rule: $Q(Gd \rightarrow Ni) + Q(Ni \rightarrow GB) = Q(Gd \rightarrow GB)$. Similar behavior as in Figs. 4-5 was observed for the Cu-doped samples.

## Energy level scheme for indium solutes.

Thermal changes in site populations can be understood in terms of two- or three-level quantum systems using the naïve interpretation that experimental values of $Q$ equal differences in the enthalpies of solutes on the various sites. Level schemes for In-probes constructed from the values of $Q$ in Fig. 4 and 5 are shown in Fig. 6. Separate diagrams are shown for the Gd-poor and Gd-rich samples.

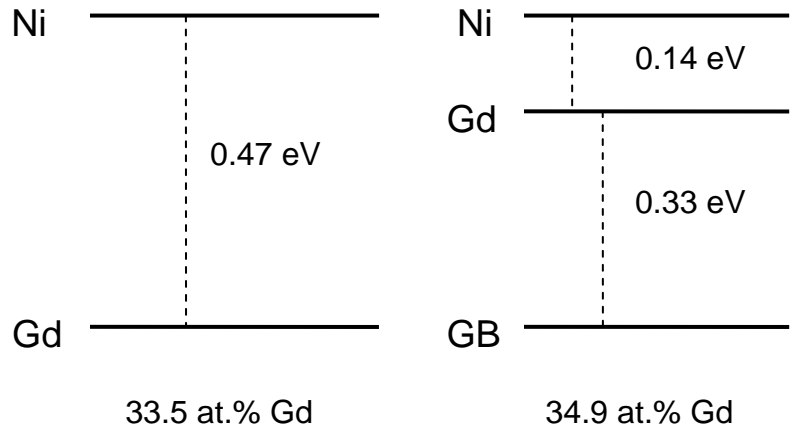

Figure 6. Energy level scheme of In probes on Ni-, Gd- and GB-sites in GdNi2 samples having 33.5 at.% Gd (left) and 34.9 at.% Gd (right). The enthalpy scales are not aligned, left and right. (The equal enthalpy ranges for the two samples is an accidental coincidence.)

One obvious difference between these schemes is the presence of grain-boundary sites at low enthalpy for the more Gd-rich sample. This implies a difference in composition of grain boundaries in the two samples. The 34.9 at.% Gd composition almost certainly exceeds the phase boundary composition, which is probably close to the stoichiometric composition. Excess Gd might form small volume fractions of a more Gd-rich phase, which were not detected in the spectra, or else might segregate at the phase boundaries, making enriched Gd layers. According to the Miedema empirical model for enthalpies of alloys, the interfacial enthalpy of In-atoms dissolved in La (taken to represent Gd) is -178 kJ/mole, much greater than in Ni, -42 kJ/mole [7]. This suggests that there is a strong driving force for In-solutes to transfer to sites within the Gd-enriched grain boundaries. However, similar GB-type sites were not observed in measurements on isostructural GdAl$_2$ [3], either pointing to a difference in GB structures between the two phases, or indicating that none of the samples studied there was actually Gd-rich. The most plausible explanation is that In-probes in Gd-rich GdNi$_2$ localize in grain boundaries that have been enriched in Gd.

The other major difference is the large change in transfer enthalpy between Gd and Ni sites in Gd-poor and Gd-rich samples, from 0.47 to 0.14 eV. Such a large change is inconsistent with the naïve view that fitted values of $Q$ using Eq. 2 are equal to the difference in enthalpies of probe atoms on the two sites. Site-enthalpies should be independent of composition since they depend on the local arrangements of atoms around the probes.

In way of explanation, an improved form of Eq. 2 takes into account the degeneracies of the levels:

$$\frac{f_2}{f_1} \cong \frac{g_2(T)}{g_1(T)} \exp(\Delta S/k_B) \exp(Q/k_B T) \approx \exp(\Delta S/k_B) \exp(Q'/k_B T). \quad (3)$$

In a parallel study of $GdAl_2(Ag)$ [8], a large change in the transfer enthalpy was also observed and can be similarly explained by strongly temperature-dependent degeneracies. A possible explanation of the change observed in the present study is as follows. With increasing temperature, Gd-atoms (and In-atoms) "evaporate" (or escape) from the Gd-rich GBs and partially fill available sites on the Ni-sublattice with Gd-atoms. This reduces the number of remaining sites that can accommodate In-solutes (that is, $g_{Ni}(T)$ decreases with increasing temperature), which in turn increases the site-fraction ratio $f_{Gd}/f_{Ni}$. The temperature dependences of the $g(T)$'s modify the overall temperature dependence, leading to fitted effective activation enthalpies $Q'$= +0.47 eV and +0.14 eV for $f_{Gd}/f_{Ni}$ that differs from the true difference in site-enthalpies $Q$ of probes on the two sites. Since there is less flow of solute atoms or excess host atoms in the undoped, Gd-poor sample, the value +0.47 eV is probably close to the site-enthalpy difference $Q$.

## Summary and conclusions.

Experiments were carried out to determine lattice locations of dilute In-solute atoms in $GdNi_2$ samples using PAC spectroscopy. $GdNi_2$ is isostructural with $GdAl_2$, which was previously studied in the same way [3]. In both phases, In-solutes are observed to transfer from Gd sites to Ni- or Al-sites with increasing temperature. In Gd-rich $GdNi_2$, an inhomogeneously broadened quadrupole interaction signal was also observed that is attributed to segregation of probe atoms to grain boundaries (GB). Such a GB site was not observed in Gd-poor $GdNi_2$, nor in $GdAl_2$. The observed enthalpy of transfer of In-solutes between Gd- and Ni-sites was 0.47 eV and 0.14 eV in Gd-poor and Gd-rich samples, respectively. Such a large change is not expected if the transfer enthalpy were simply a difference in site-enthalpies. The change is attributed to strong temperature dependences of the *degeneracies* of the Gd- and Ni-levels available for In-solutes that vary with composition. As a possible mechanism, it was proposed that an increase in temperature in the Gd-rich sample might lead to release of Gd-atoms from GBs that combine with virtual vacancies on the Ni-sublattice to make antisite atoms, thereby reducing the number of available sites for In-solutes. Such a mechanism would lead to a strong temperature dependence of the ratio of site-degeneracies that appears in Eq. 3, which combines with the unique site-enthalpy difference $Q$ to produce the effective transfer enthalpy $Q'$ that is observed. Finally, distinct GB segregation enthalpies were obtained for solutes on the Gd- and Ni-sublattices that would appear only as an average value when obtained using a macroscopic method.

## Acknowledgements.

This work was supported in part by the National Science Foundation under grant DMR 14-10159 (MMN Program).

1  Helmut Mehrer, Diffusion in Solids: Fundamentals, Methods, Materials, Diffusion-Controlled Processes, (Springer Series in Solid-State Sciences 155, Berlin, 2007).
2  P. Villars and L.D. Calvert, Pearson's Handbook of Crystallographic Data for Intermetallic Compounds, 2$^{nd}$ ed. (ASM International, Materials Park, Ohio, 1991).
3  Matthew O. Zacate and Gary S. Collins, Phyical Review B69, 174202(1-9) (2004).
4  Gary S. Collins, J. Materials Science 42, 1915-1919 (2007).
5  M. Forker, Nuclear Methods and Instrumentation 106, 121-126 (1973).
6  G. Schatz and A. Weidinger, Nuclear Condensed-Matter Physics, (John Wiley, New York, 1996).
7  See, e.g., H. Bakker, Enthalpies in Alloys: Miedema's Semi-Empirical Model, (Trans Tech Publications, Materials Science Foundations, vol. 1, Switzerland, 1998), Table A3, p. 71-72.
8  Debashis Banerjee, Ryan Murray and Gary S. Collins, "Solute-solute interactions in intermetallic compounds", (companion paper to this conference; contribution #40).